\documentclass[aps,prb,showpacs,reprint,superscriptaddress]{revtex4-1}

\usepackage{graphicx}
\usepackage{color}
\usepackage{amsmath}
\usepackage{bbm}
\usepackage{amssymb}

\usepackage{dsfont}

\usepackage[utf8]{inputenc}	
\usepackage[T1]{fontenc}

\input epsf

\begin{document}

\title{Variational Monte-Carlo Approach for Hubbard Model Applied to Twisted Bilayer WSe$_2$ at Half-Filling}
\author{A. Biborski}
\email{andrzej.biborski@agh.edu.pl}
\affiliation{Academic Centre for Materials and Nanotechnology, AGH University of Krakow, Al. Mickiewicza 30, 30-059 Krakow,
Poland}
\author{P. W\'ojcik}
\affiliation{AGH University of Krakow, Faculty of Physics and Applied Computer Science,30-059 Krakow, Al. Mickiewicza 30, 30-059 Krakow, Poland}
\author{M. Zegrodnik}
\affiliation{Academic Centre for Materials and Nanotechnology, AGH University of Krakow, Al. Mickiewicza 30, 30-059 Krakow,
Poland}

\begin{abstract}
We consider an effective Hubbard model with spin- and direction-dependent complex hoppings $t$, applied to twisted homobilayer WSe$_2$ using a variational Monte Carlo approach. The electronic correlations are taken into account by applying the Gutzwiller on-site correlator as well as long-range Jastrow correlators subjected to noninteracting part being of Pfaffian form. Our analysis shows the emergence of Mott insulating state at critical value of Hubbard interaction $U_{c1}\approx 6.5|t|\div7|t|$ estimated by extrapolating the density-density equal-time two-particle Green's functions. The signatures of an intermediate insulating phase between $U_{c1}$ and $U_{c2}\approx9.5|t|\div10|t|$ are also discussed. Furthermore, we report the formation of the $120^{\circ}$ in-plane Néel state indicated by the detailed analysis of the spin-spin correlation functions. As shown, switching between antiferromagnetic phases characterized by opposite chirality could be experimentally realized by the change of perpendicular electric field. In proper range of electric fields also a transition to in-plane ferromagnetic state appears.
\end{abstract}

\maketitle

\section{Introduction}
The role of electronic correlations in a variety of condensed matter systems remains mysterious in many cases. In particular, the mechanism leading to the emergence of unconventional superconductivity or the formation of long-range charge and spin-ordered states is not fully recognized\cite{Dagotto2005}. In this view, two-dimensional moir\'e superlattices\cite{XIAO20201142}, in which the Fermi level\cite{Cao2018_1,Cao2018_2,Wang2020,Ghiotto2021} as well as the electron-electron interaction strength can be relatively easily tuned\cite{Wang2020,Ghiotto2021}, are considered as a promising platform for a better understanding of correlation-driven phenomena. 

The moir\'e structures are typically fabricated by twisting two or more layers of a given material, which eventually leads to the formation of narrow bands\cite{Balents2020,Hongyuan2021,Xu_2022}. In such situation, the role of electronic interactions is believed to be enhanced\cite{Haining2020}. An archetypic example of this scenario is realized in twisted bilayer graphene (TBG), in which for certain twist angles (the so-called \emph{ magic angles}) insulating (Mott) and unconventional superconducting states have been reported\cite{Cao2018_1,Cao2018_2,Zhang_2021_prl,Chen2019,Shen2020}. However, experimentally realized moir\'e superlattices are not limited to TBG, or more generally to twisted multilayers of graphene\cite{Zhang_2021_prl, Chen2019, Shen2020,Fidrysiak2023}. As recently reported, the twisted homobilayers and heterobilayers\cite{Haining2020} based on transition metal dichalcogenides (TMD) also exhibit phenomenona driven by electronic correlations together with signatures of the superconducting state\cite{Wang2020,Ghiotto2021,Huang2021}. In these systems, the appearance of narrow bands is less sensitive to changes in the twist angle between the layers\cite{Wang2020,Ghiotto2021,Haining2020} when compared to the graphene-based structures. Moreover, for the heterobilayer case, this effect appears even in the absence of the twist angle due to the mismatch of the lattice constants of the two monoatomic layers. Another important feature of TMDs is the presence of spin-valley locking\cite{Xiao2012,Bawden2016,Wang2020ACS,Tao_2023}. For heterobilayers such as WSe$_2$/MoSe$_2$ this effect can be considered as intrinistic and based on the breaking of the inversion symmetry\cite{Haining2020}, while in homobilayers it can be tuned by the perpendicular external electric field (displacement field)\cite{Haining2020,Tao_2023}.

As recently shown, the bilayer moir\'e TMDs can be efficiently described by a Hubbard-type Hamiltonian on a triangular lattice with spin-dependent complex hoppings\cite{Haining2020,Rademaker2022}. Such an approach incorporates both the valley-dependent spin-splitting as well as the physics of strong electronic correlations. Based on semiclassical arguments applied to the Hubbard model mapped to the Heisenberg Hamiltonian\cite{Haining2020}, it has been suggested that a $120^o$ in-plane Néel ordering should appear in the twisted WSe$_2$ homobilayer. A more detailed analysis of the magnetically ordered states has been carried out with the use of Hartree-Fock and cluster dynamical mean-field theory by J. Zhang et al.\cite{Zang2021,Zang2022}. Also, recently an extended Hubbard model supplemented with an intersite Coulomb repulsion has been studied from the point of view of charge and spin ordered states in the WSe$_2$/WS$_2$ heterostructures\cite{Rademaker2023, Rademaker_2023_2}. 

It should be noted that the elaborated Hubbard model with spin-dependent hopping phase can also be regarded as interesting \emph{per se}. As the Hubbard model, despite its simplicity, is supposed to predict an astonishingly rich phase diagram\cite{Arovas2022}, there are also still open questions regarding its realization in the triangular lattice. In particular, the formation of the quantum spin liquid phase (QSL)\cite{Balents2010,Shirakawa,Szasz,Chen2022,Zampronio2023} preceding the antiferromagnetic insulator (AFI) with an increasing value of $U$ remains enigmatic. In this spirit, the reduction of degrees of freedom by imposing spin-valley locking in free-particle terms of the Hamiltonian may help to shed light on this issue from both the experimental and theoretical points of view.

In our previous report we analyzed the stability of mixed singlet-triplet paired state in tWSe$_2$ by applying the Gutzwiller approach to the so-called $t-J-U$ Hamiltonian\cite{Zegrodnik2023}. Here, we turn to the study of Mott-insulating and magnetically ordered states (antiferro- and ferro-magnetic) within the original Hubbard picture by using the Variational Monte Carlo (VMC) method. We focus on the half-filled band case and analyze the evolution of the ground state of the system from weak to strong correlations. According to our analysis, robust in-plane antiferromagnetic ordering appears in a relatively wide range of Hubbard $U$. We also study the influence of the complex phase of the hoppings, which can be tuned experimentally by the displacement field, and determines the magnitude of the valley-dependent spin-splitting. Interestingly, our results explicitly show the possibility of switching between the two antiferromagnetic states with different chirality as well as between antiferro- and ferro-magnetic alignment. 

The paper is organized as follows. First, we briefly describe the employed model, providing also the sketch of the Variational Monte Carlo (VMC) approach which has been utilized. Next, we study the metal insulator-transiton (MIT), based on the equal-time one- and two-particle Green's functions. Finally, we investigate in detail the magnetic properties of the system by means of analysis performed in both real and momentum space.


\section{Model and method}

\begin{figure}
 \centering
 \includegraphics[width=0.4\textwidth]{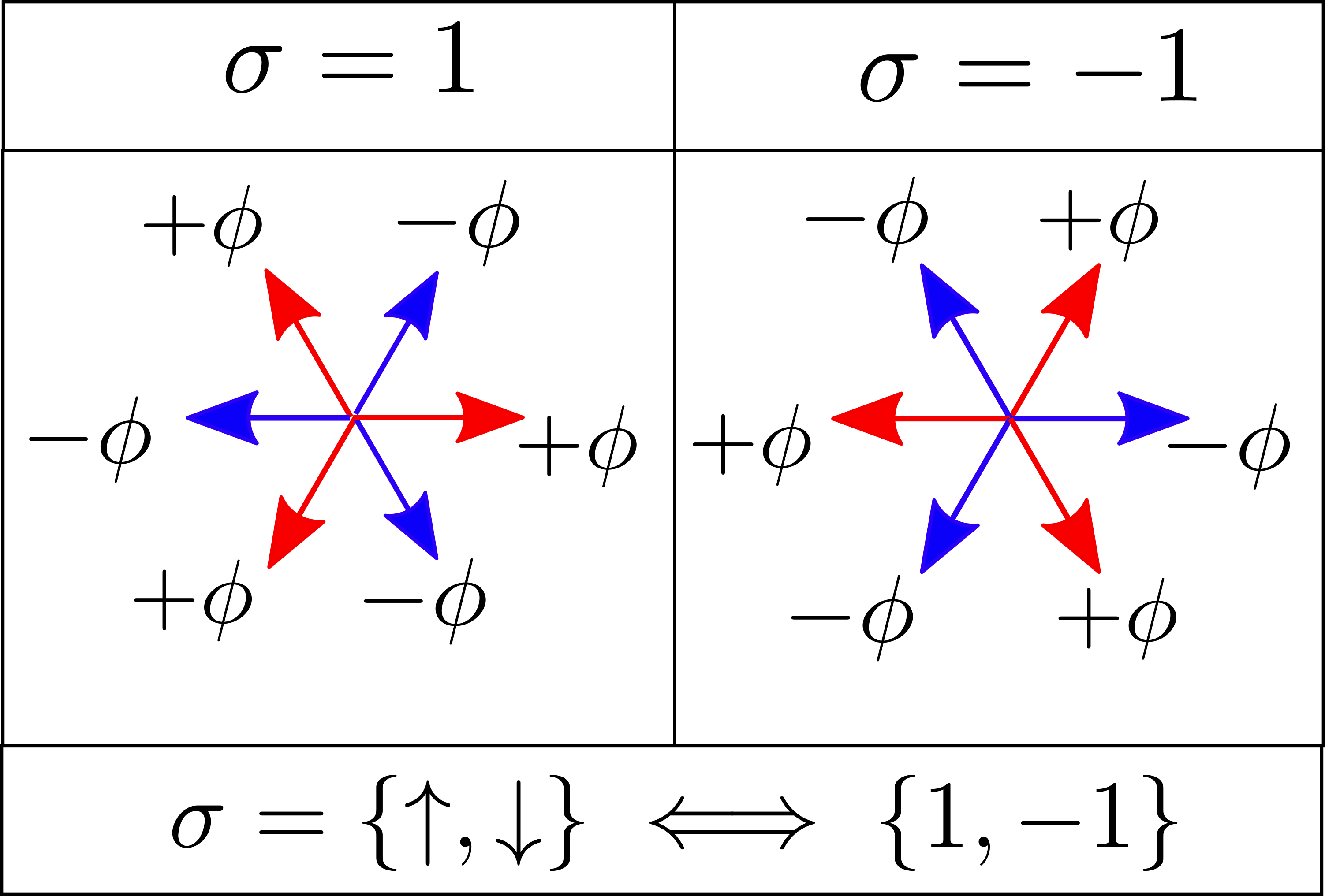}
 \caption{Schematic representation of hopping phases according to spin and direction.} 
 \label{fig:hoppings}
\end{figure}

We consider the minimal model describing moir\'e superlattice of WSe$_{2}$ tTMD as derived
by Haining and Das Sarma\cite{Haining2020} from the continuous approach, i.e.,
\begin{align}
\label{eq:Hamitlontian}
\Hat{H}=\sum_{\langle ij\rangle}\sum_{\sigma}|t|e^{i\phi_{ij}\sigma}\hat{a}_{i\sigma}^{\dagger}\hat{a}_{j\sigma}+U\sum_{i}\hat{n}_{i\uparrow}\hat{n}_{i\downarrow},
\end{align}
where $\hat{a}_{i\sigma}^{\dagger}$($\hat{a}_{i\sigma}$) is the standard creation (anihilation) operator of electron at site $i$ with spin $\sigma=\{1,-1\}$, while $\hat{n}_{i\sigma}\equiv\hat{a}_{i\sigma}^{\dagger}\hat{a}_{i\sigma}$ is the carrier occupation operator. The sum of $\langle ij \rangle$ in the kinetic term denotes that only the nearest neighboring (nn) sites on the triangular lattice are taken into account, since the more distant sites are believed to be of significantly lower amplitude and therefore play a marginal role. As the hermicity of the Hamiltonian must be conserved, $\phi_{ij}=-\phi_{ji}$.  The sketch of the phase of the carriers' hoppings for different spins and different directions is provided in Fig.~\ref{fig:hoppings} where we set $\phi_{ij}=\phi$ for the right-hand side nearest neighbor. It should be noted that such structure of the complex hoppings incorporates the valley-dependent spin-splitting induced by an effective spin-orbit interaction. The value of $\phi$ can be tuned by the displacement field across the bilayer, which is generated by the top and bottom gates in the real experimental setup\cite{Haining2020,Wang2020}. The second term of our Hamiltonian represents the on-site Coulomb repulsion with the value of the Hubbard $U$ parameter depending mainly on the twist angle.

To take into account electron-electron correlation effects resulting from significant magnitude of the Hubbard $U$ in Eq.~(\ref{eq:Hamitlontian}) we employ the VMC\cite{Becca,MISAWA2019} approach, which is based on the variational formulation, i.e.,
\begin{align}
\label{eq:variational_statement}
E_{G}\leq E_T\equiv\frac{\langle\Psi_T|\mathcal{\hat{H}}| \Psi_T\rangle}{\langle\Psi_T|\Psi_T\rangle},    
\end{align}
where $E_G$ is the unknown ground-state energy to be estimated by the trial energy $E_T$. The latter is given as the expected value of the Hamiltonian calculated with respect to the variational state $|\Psi_T\rangle$, which is the subject of optimization. The trial, parametrized state, $|\Psi_T\rangle$ needs to be \emph{a priori} proposed and is taken as
\begin{align}
    |\Psi_T\rangle =  \mathcal{\hat P}_G(\{g_i\})\mathcal{\hat P}_J(\lambda_{ij})\mathcal{\hat{L}}_{S^z}\mathcal{\hat{L}}_{N_e} |\Psi_0\rangle,
\end{align}
where $\mathcal{\hat{P}}_{G}(\{g_i\})=e^{-\sum_{i}g_i\hat{n}_{i\uparrow}\hat{n}_{i\downarrow}}$ is the Gutzwiller type local (on-site) correlator projecting out the double occupied sites, whereas $\mathcal{\hat{P}}_{J}(\{\lambda_{ij}\})=e^{-\sum_{i,j}\lambda_{ij}\hat{n}_{i}\hat{n}_{j}}$ is the Jastrow correlator accounting for the non-local (off-site) density-density correlations. The operators $\mathcal{\hat{L}}_S^{z}$ and $\mathcal{\hat{L}}_N^{e}$ project the trial many body state onto the Fock subspace in which the total spin component in the $\mathbf{z}$ axis is set to zero, and the number of particles is set to $N_e$, respectively. The noncorrelated part $|\Psi_0\rangle$ is given in the so-called Pfaffian form\cite{bouchaud1988,Bajdich,Becca,MISAWA2019}, 
\begin{align}
    \mathcal{\hat{L}}_N^{e}|\Psi_0\rangle=\Bigg[\sum_{i,j\sigma,\sigma'}F_{i,j}^{\sigma \sigma'}\hat{c}_{i,\sigma}^{\dagger}\hat{c}_{j,\sigma'}^{\dagger}\Bigg]^{N_e/2}|0\rangle,
\end{align}
thus it contains both $\hat{c}_{i,\sigma}^{\dagger}\hat{c}_{j,\sigma}^{\dagger}$(parallel-spins) and $\hat{c}_{i,\sigma}^{\dagger}\hat{c}_{j,\bar{\sigma}}^{\dagger}$(antiparallel-spins) pairings. 

\begin{figure}
    \includegraphics[width=0.4\textwidth]{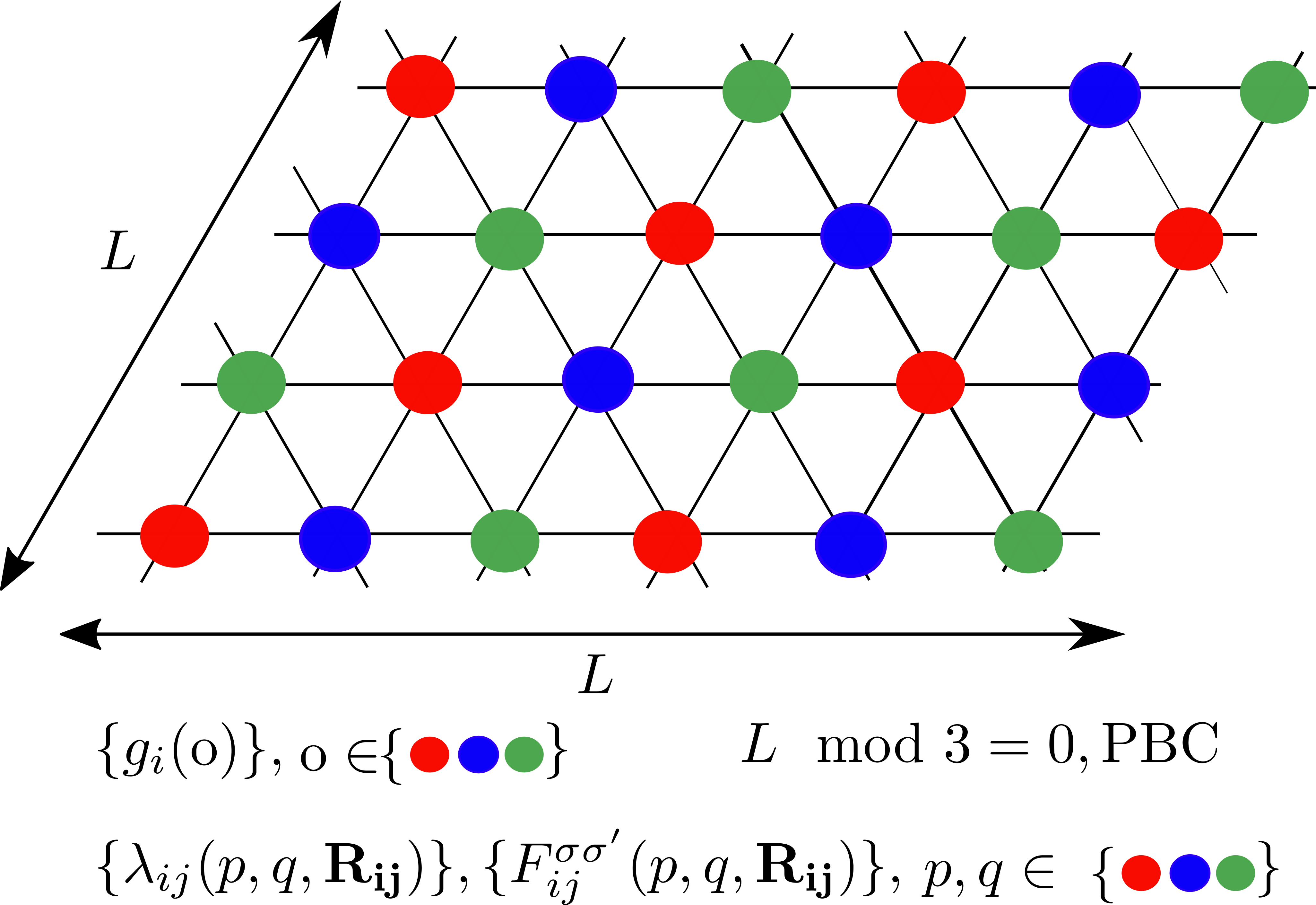}
    \caption{Structure of variational parameters used for the formulation of $|\Psi_T\rangle$.}
    \label{fig:ansatz}
\end{figure}

Eventually, $\{g_i\},\{\lambda_{ij}\}$ and $\{F_{i,j}^{\sigma\bar{\sigma}}\}$ span the space of variational parameters to be optimized. 
The Hamiltonian given in Eq.~(\ref{eq:Hamitlontian}) at large $U/|t|$ can be mapped onto the anisotropic Heisenberg model supplemented with the effective Dzyaloshinskii-Moriya term. By using semiclassical arguments, it has been proposed in Ref. \onlinecite{Haining2020} that such a Hamiltonian should lead to a series of $\phi$-dependent in-plane spin ordered states allowing for switching between $120^{\circ}$ AF and FM states. Since we intend to examine the original Hubbard model without the explicit inclusion of the exchange term, we need to apply the properly constructed variational \emph{ ansatz} $|\Psi_T\rangle$. Reconstruction of $120^{\circ}$ AF in-plane ordering requires the provision of at least a three-sublattice structure for the variational state. Also, since we subject the supercell of size $L\times L$ to the periodic bounduary condition, the $L\mod3=0$ requirement must be met to avoid possible spin frustration in the $\mathbf{x}-\mathbf{y}$ plane. All subsets of variational parameters exhibit the aforementioned three-sublattice structure as shown in Fig.~\ref{fig:ansatz}. However, intersite variational parameters, e.g. $\lambda_{ij}$ (which can be in general nonzero for all possible pairs $(i,j)$), are not presumed to be equal for given $|\mathbf{R_{ij}}|$, that is, they are allowed to be different with respect to the direction of vector $\mathbf{R}_{ij}$ pointing from $i$th to $j$th lattice site. Also, while $\lambda_{ij}=\lambda_{ji}$ and $F_{ij}^{\sigma\sigma}=F_{ji}^{\sigma\sigma}$ the antiparallel spin-spin parameters may differ with respect to the transpose of spatial indicies. 
Finally, the three-sublattice structure is the only symmetry requirement provided \emph{ad hoc} in $|\Psi_T\rangle$, and we do not impose any other spatial restrictions. Thus, remaining degrees of freedom are relaxed and other symmetries of the Hamiltonian can be broken in the resulting ground-state solution.   
The right-hand side of Eq.~(\ref{eq:variational_statement}) is optimized by applying the \emph{ stochastic reconfiguration method}\cite{Becca}, where both $F_{ij}^{\sigma\sigma}$ and $F_{ij}^{\sigma\bar{\sigma}}$ are allowed to be complex numbers. In all VMC calculations, we used the mVMC software provided by Misawa et al.\cite{MISAWA2019}.


\section{Results}

In this Section, we first analyze the emergance of the Mott phase, i.e. the transition between metallic and correlation-driven gapped state. Subsequently, we focus on the spin-ordered phases. Unless stated otherwise, we take $t=-6.95\pm 5.03i$ (meV), which is the value reported by Wang et al.\cite{Wang2020}, corresponding to the twist angle $\Theta=5.08^{\circ}$ and the displacement field $0.45$ (V/nm). These parameters are estimated to reproduce an abrupt increase in resistivity measured for tWSe$_2$ homobilayer at half filling\cite{Wang2020}. Note also that $\arg(t)\approx\pm4\pi/5$, thus it should reconstruct the in-plane $120^{\circ}-AF$ order as predicted by Pan et al.\cite{Haining2020}.
In the last part of this Section, we study the influence of phase $\phi$ on the magnetic properties of the system in the strongly correlated regime.

\subsection{Mott phase formation}

\begin{figure}
    \includegraphics[width=0.5\textwidth]{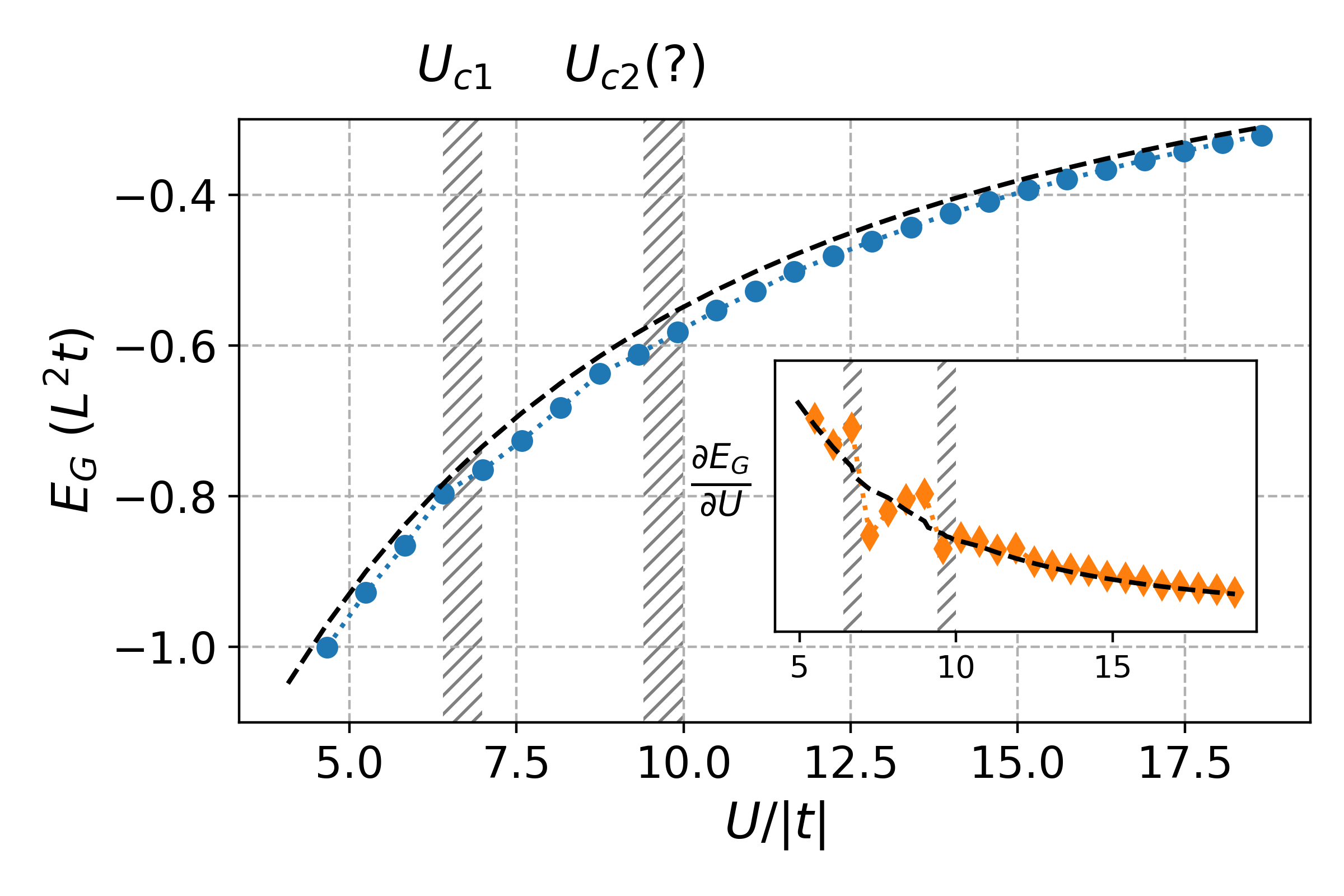}
    \caption{Ground state energy $E_G$ obtained for the wide range of $U/|t|$ by means of VMC  approach (symbols and dotted line) and unrestricted Hartree-Fock (dashed black line). The two vertical dashed regions indicate values of $U$ for which an anomalous behavior of the energy curve appears. This becomes more evident in the plot of $\partial E_G/\partial U$ provided in the inset.}
    \label{fig:energy}
\end{figure}

\begin{figure}
    \includegraphics[width=0.5\textwidth]{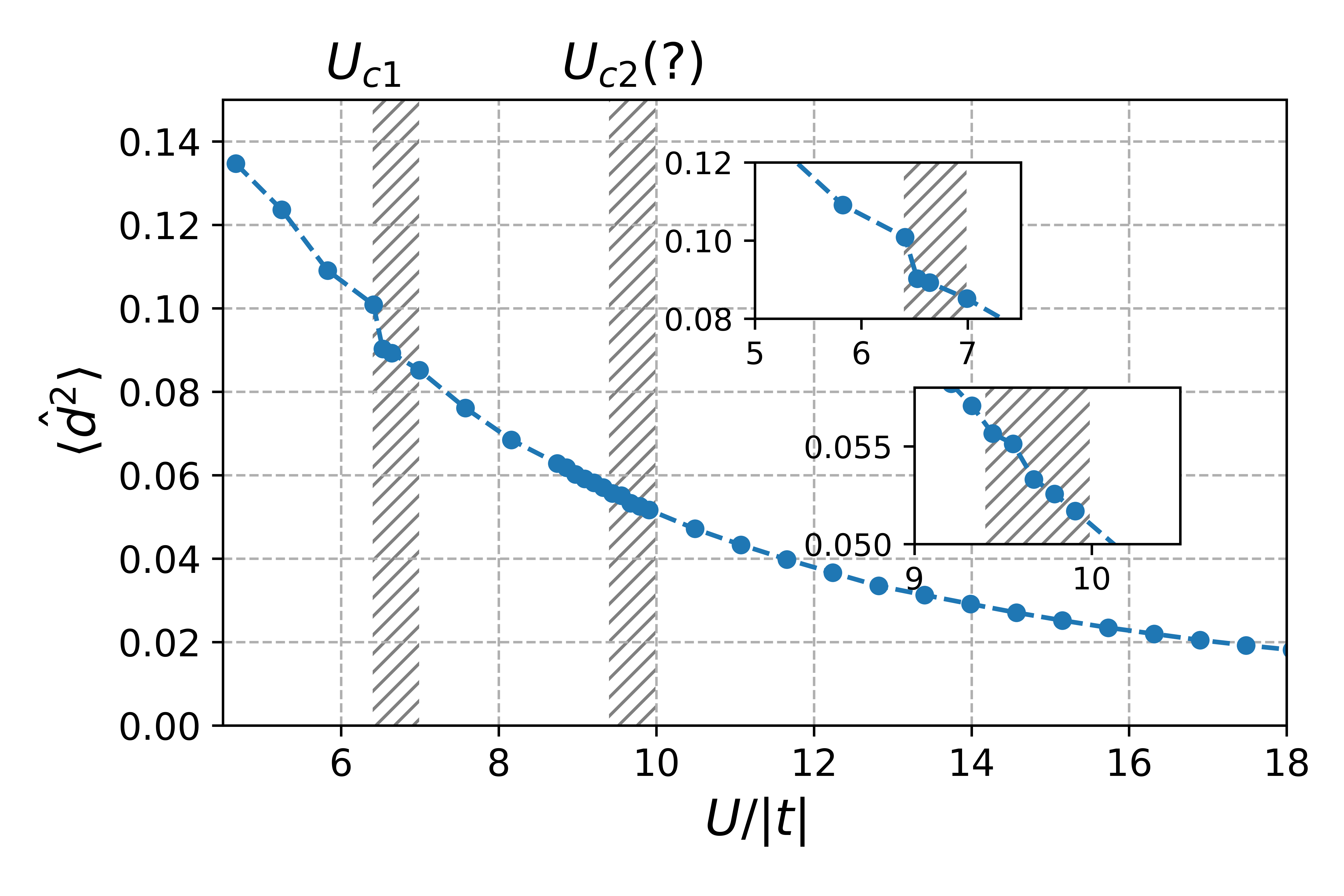}
    \caption{Double occupancy $\langle \hat{d}^2\rangle$ as a function of $U$. An abrupt decrease of double occupancies is clearly visible in vicinity of $U_{c1}$, whereas for $U_{c2}$ it is much less pronounced. }
    \label{fig:d2}
\end{figure}

First, we analyze the dependence of the total energy of the system per site for the supercell comprising the $L\times L=12\times 12$ lattice sites at half-filling, i.e. $N_e=L^2$. As can be seen from the plot presented in Fig.~\ref{fig:energy} the mean field treatment (unrestricted Hartree-Fock in our case) does not provide any evidence of anomalous energy behavior as a function of $U$. On the contrary, the application of the VMC approach leads to a discontinuity in $\partial E_G/\partial U$ at $U_{c1}\approx6.5|t|$ and $U_{c2}\approx 9.5|t|$, with the latter being less pronounced. Such behaviour, cannot yet be understood as a clear evidence of the Mott transition, therefore, we have analyzed the average value of doubly occupied sites defined as
\begin{align}
    \langle \hat{d}^2\rangle\equiv\frac{1}{L^2}\sum_{i}\langle\hat{n}_{i\uparrow}\hat{n}_{i\downarrow}\rangle.
    \label{eq:d}
\end{align}
In Fig.~\ref{fig:d2} we show that in the vicinity of  previously identified $U_{c1}\approx6.5 |t|$ the value of $\langle\hat{d}\rangle$ abruptly decreases indicating the phase transition to the Mott insulating phase\cite{Shirakawa,Yoshioka}. Similar behavior is also present near $U_{c2}$, however, in this case the effect is much less pronounced and is hardly distinguishable from the uncertainties inherent in the VMC approach. 

\begin{figure}
    \includegraphics[width=0.4\textwidth]{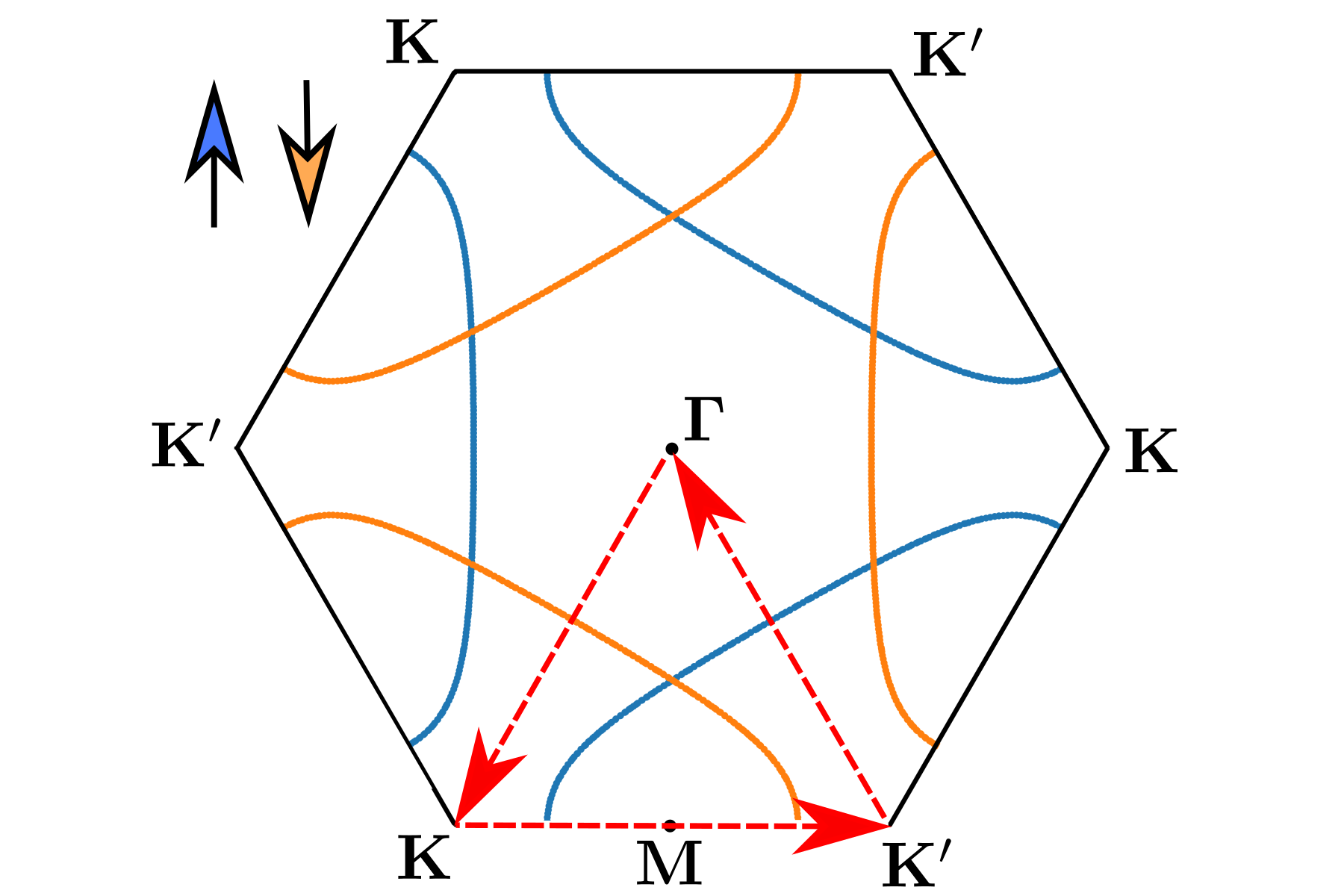}
    \caption{The Fermi surface (solid lines) for the spin-splitted bare-band ($U=0$). The colors distinguish spin-up and spin-down sub-bands. The red dashed arrows indicate the trajectory, containing the high symmetry points, along which the momentum resolved electron occupation number has been calculated and is provided in Fig. \ref{fig:nk}.}
    \label{fig:bz}
\end{figure}

Next, we turn to the analysis of the correlation-induced insulating state from the point of view of the momentum-resolved electron occupancy. Here, we investigate the quantity defined as
\begin{align}
\langle \hat{n}_{\mathbf{q}\sigma}\rangle\equiv\frac{1}{L^2}\sum_{ij}\exp{(-i\mathbf{q}\cdot\mathbf{R}_{ij})}\langle\hat{a}_{i\sigma}^{\dagger}\hat{a}_{j\sigma}\rangle,
\end{align}
where $\mathbf{q}$ is a vector in the momentum space. Clear discontinuities in $\langle \hat{n}_{\mathbf{q}\sigma}\rangle$ are likely to appear when the Fermi surface is crossed as we go along the trajectory inside the Brillouin zone (see arrows in Fig.~\ref{fig:bz}) for the case of non- or weakly correlated regimes. In contrast, when the quasiparticle spectrum is largely renormalized by the strength of interactions, the discontinuity should become more and more suppressed, while its height can be considered as weight-measuring quasiparticle coherence (quasiparticles dressed in interactions)\cite{Spalek2022}.

\begin{figure}
    \includegraphics[width=0.5\textwidth]{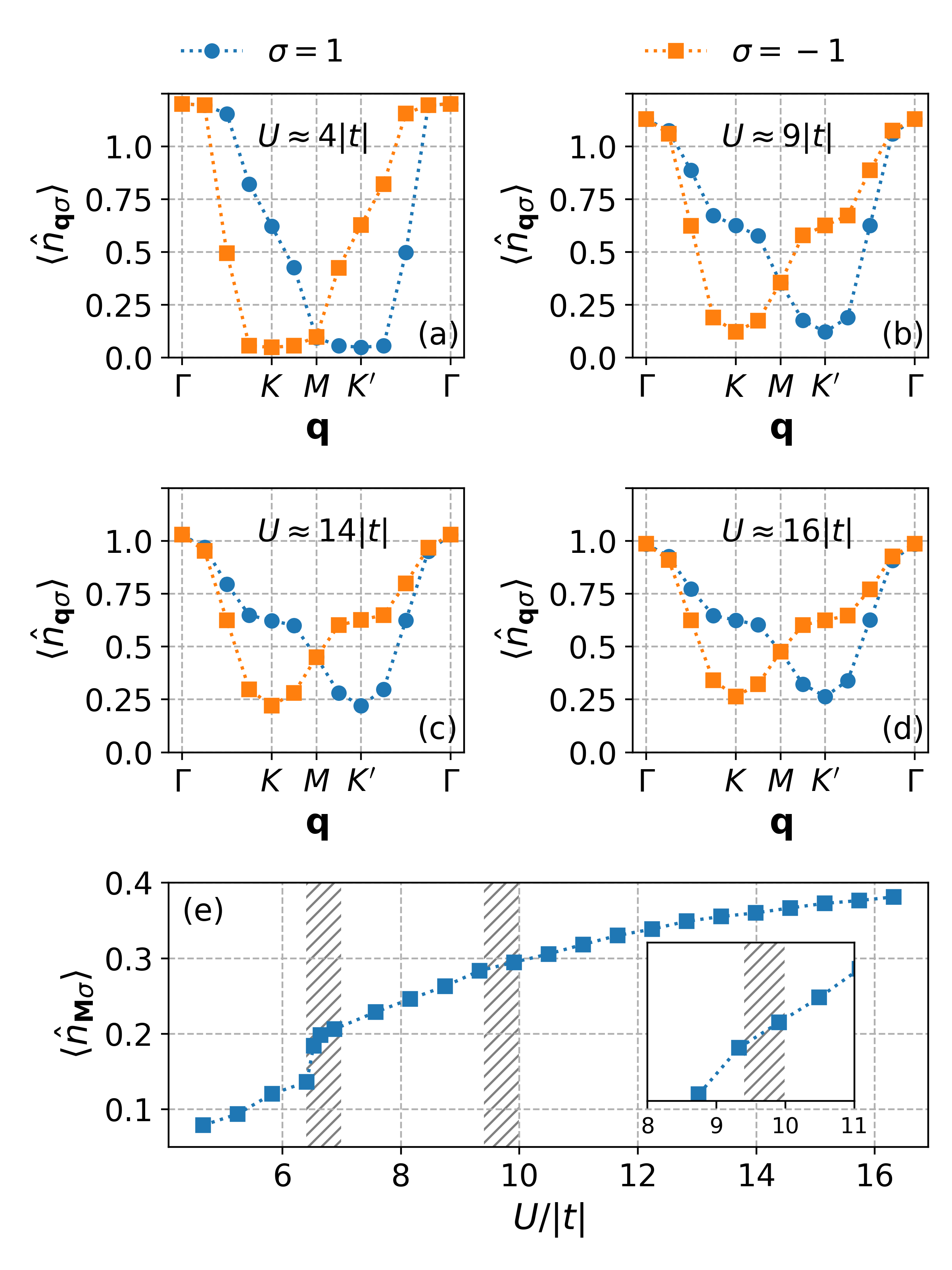}
    \caption{(a-d) Momentum resolved occupation number for both spin directions $\sigma$, along the path defined in Fig. \ref{fig:bz} for different values of $U$. (e) Fourier transform of electron occupancy at the high symmetry point $\mathbf{q}=\mathbf{M}$.}
    \label{fig:nk}
\end{figure}

As shown in Fig.~\ref{fig:nk} the quantitative change in $\langle \hat{n}_{\mathbf{q}\sigma}\rangle$ appears with increasing amplitude of the Hubbard interaction. Namely, for $U\approx 4|t|$ the relatively clear discontuities can be identified when the Fermi surface is crossed along the $\mathbf{K}-\mathbf{\Gamma}$ and $\mathbf{\Gamma}-\mathbf{K'}$ vectors for the up and down spins, respectively [see Fig.~\ref{fig:nk}(a)]. As $U$ increases, the states above the Fermi level (in a renormalized picture) become also occupied [Figs.~\ref{fig:nk}(b-c)] due to interactions, as expected. In particular, the states related to the high symmetry point $\mathbf{M}$ in the $\mathbf{q}$ space are initially almost empty, but for the larger values of $U$ they start to participate in the occupation scheme renormalized by electron-electron interactions. 

To visualize this effect, we plot $\langle \hat{n}_{\mathbf{q}\sigma}\rangle$ for $\mathbf{q}=\mathbf{M}$ (note that for this high symmetry point $\langle \hat{n}_{\mathbf{q}\uparrow}\rangle=\langle \hat{n}_{\mathbf{q}\downarrow}\rangle$) as a function of $U$, in Fig.~\ref{fig:nk}(e). In addition to the gradually increasing occupancy of the states above the Fermi level with increasing $U$ which is driven by \emph{dressing} the quasiparticles with interactions, we observe an abrupt jump in $\langle \hat{n}_{\mathbf{M}\sigma}\rangle$ at $U_{c1}$, indicating a critical behavior corresponding to the transition to the Mott insulating state. Such a jump is not clear at $U_{c2}$.

For the sake of completeness, we have analyzed yet another quantity which can indicate the transition to the Mott insulating state. Namely, the Fourier transform of the density-density correlation function, i.e.,
\begin{align}
    \mathcal{N}(\mathbf{q})\equiv\frac{1}{L^2}\Big\langle\sum_{i,j}e^{-i\mathbf{q}\cdot(\mathbf{R}_i-\mathbf{R}_j)}\hat{n}_i\hat{n}_j\Big\rangle, 
\end{align}
where $\hat{n}_i=\hat{n}_{i\uparrow}+\hat{n}_{i\downarrow}$. As discussed in Refs. \onlinecite{Capello2005,Tocchio2011,Tocchio2020}, $\lim_{\mathbf{q}\rightarrow0} q^2/{\mathcal{N}(\mathbf{q})} \sim \Delta_G$, where $\Delta_G$ is the magnitude of the gap driven by electron-electron interactions. 

\begin{figure}
    \includegraphics[width=0.5\textwidth]{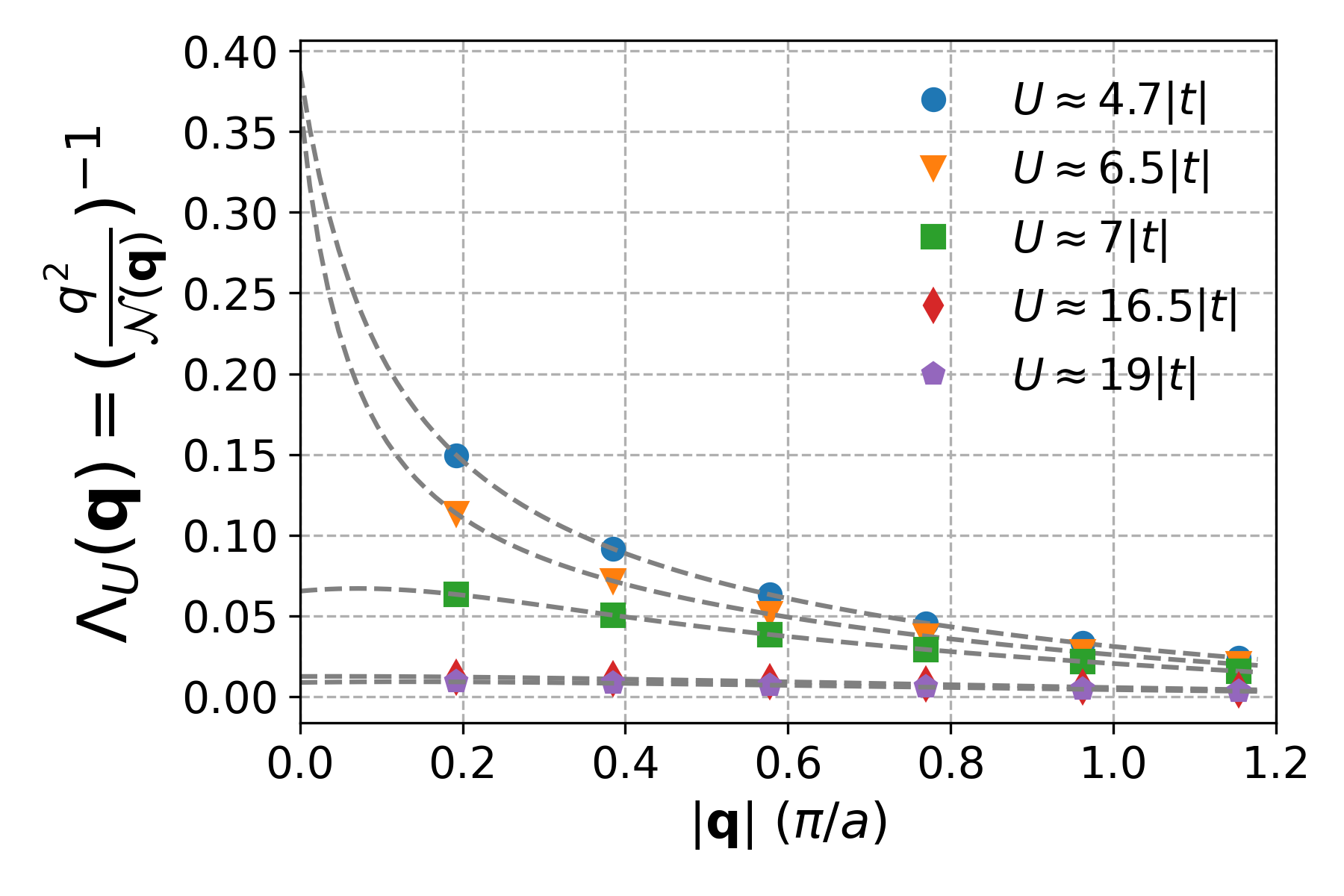}
    \caption{The inverse of $q^{2}/\mathcal{N}(\mathbf{q})\equiv \Lambda_U(\mathbf{q})$ as a function of $|\mathbf{q}|\in \mathbf{\Gamma}-\mathbf{K}$ for the selected values of $U$. The dashed lines refer to the polynomial (5th order) fits.} 
    \label{fig:nnvsu}
\end{figure}

In Fig.~\ref{fig:nnvsu} we present $\Lambda_U(\mathbf{q})\equiv\big[q^2/\mathcal{N}(\mathbf{q})\big]^{-1}$ for the selected representative values of $U$. Note that we plot the inverse, since the transition associated with $U_{c1}$ is considerably more visible in such a case. As shown in Fig.~\ref{fig:nnvsu} for relatively low values of $U$ (weakly to moderately correlated), we observe a substantial increase in $\Lambda$ with $|\mathbf{q}|\rightarrow 0$ which indicates that the gap is closed there. However, in the large $U$ regime we observe a significant change of $\Lambda_U(\mathbf{q})$ behavior, which points to the creation of the insulating state. Of course, since our supercell is of \emph{finite} size, the $q=0$ limit cannot be achieved because the number of points in the $\mathbf{q}$-space equals the number of sites. In our analysis, we consider the trajectory $\mathbf{K}\rightarrow\mathbf{\Gamma}$ which contains six equidistant $\mathbf{q}$ points. To overcome the issue related to the finite system, we performed fits of the fifth degree polynomial for $1/\Lambda_{U}(\mathbf{q})$ (marked as dashed lines in Fig.~\ref{fig:nnvsu}). In this manner, we can extract the behavior of $\Lambda_U(\mathbf{q})$ in the vicinity of $\mathbf{\Gamma}$ point.

\begin{figure}
    \includegraphics[width=0.5\textwidth]{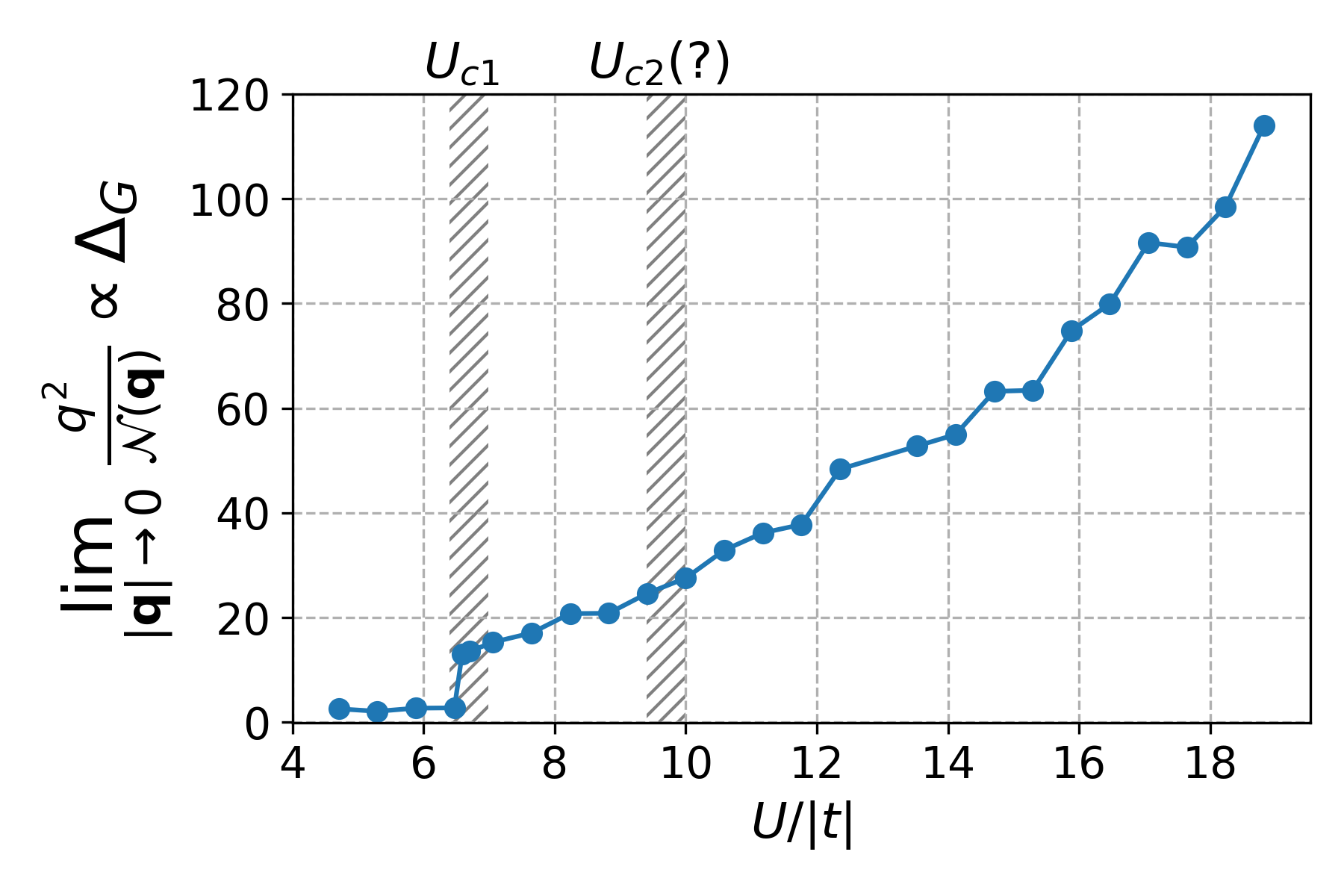}
    \caption{The estimate of  $\lim_{\mathbf{q}\rightarrow0}q^{2}/\mathcal{N}(\mathbf{q})$ as a function of interaction magnitude $U$. This quantity is proportional to the magnitude of the correlation-induced gap~\cite{Capello2005}.} 
    \label{fig:gap_estimation}
\end{figure}

In Fig.~\ref{fig:gap_estimation} we show $1/\Lambda_{U}(q=0)\propto\Delta_G$ as a function of $U$, obtained by computing the aforementioned (fitted) polynomials at $q=0$. The result is consistent with the analysis of total energy, double occupancies $\langle\hat{d}^{2}\rangle$, and $\langle\hat{n}_{\mathbf{M}\sigma}\rangle$. Namely, at $U\approx U_{c1}$ we observe an abrupt jump in $1/\Lambda_{U}(q=0)$ pointing to the opening of the gap. The uncertainties observed for the higher values of $U$ ($\gtrsim 12|t|$) discriminate the realiable inspection in the vicinity of $U_{c2}$. Thus, from this perspective, it remains enigmatic if the \emph{ intermediate} phase between $U_{c1}$ and $U_{c2}$ emerges or if the observed anomalies in $\partial E_G/\partial U$ and $\langle\hat{d}^2\rangle$ in $U_{c2}$ are just numerical artifacts. Nevertheless, since the existence of spin-liquid phase for $U$ lying between $U_{c1}\approx8|t|$ and $U_{c2}\approx10|t|$ has been reported for the isotropic triangular Hubbard model treated by unbiased methods (see Ref. \onlinecite{Arovas2022} and the references therein), it is tempting to identify even weak signatures of the presence of an intermediate (possibly spin-liquid) phase in our variational picture. Thus, we intentionally also paid some attention to this aspect.

\subsection{Spin order}
In this Section we analyze in detail the magnetic properties of our system.
It should be noted that the interplay between the spin-orbit coupling, encapsulated in the noninteracting part of the Hamiltonian (Eq.~\ref{eq:Hamitlontian}), and the strong Coulomb repulsion may alter the resulting magnetic properties. Namely, at $U\gg|t|$ the original Hubbard model can be transformed to the anisotropic, $\phi$-dependent Heisenberg model supplied with the Dzyaloshinskii-Moriya term~\cite{Haining2020,Haining2022} that for certain values of $\phi$, projects out-of-plane AF order onto the $120^{\circ}$ Néel state in the $\mathbf{x}-\mathbf{y}$ plane. Therefore, here we study how to tune the magnetic properties of the system by changing $\phi$.

\begin{figure}
    \includegraphics[width=1.0\linewidth]{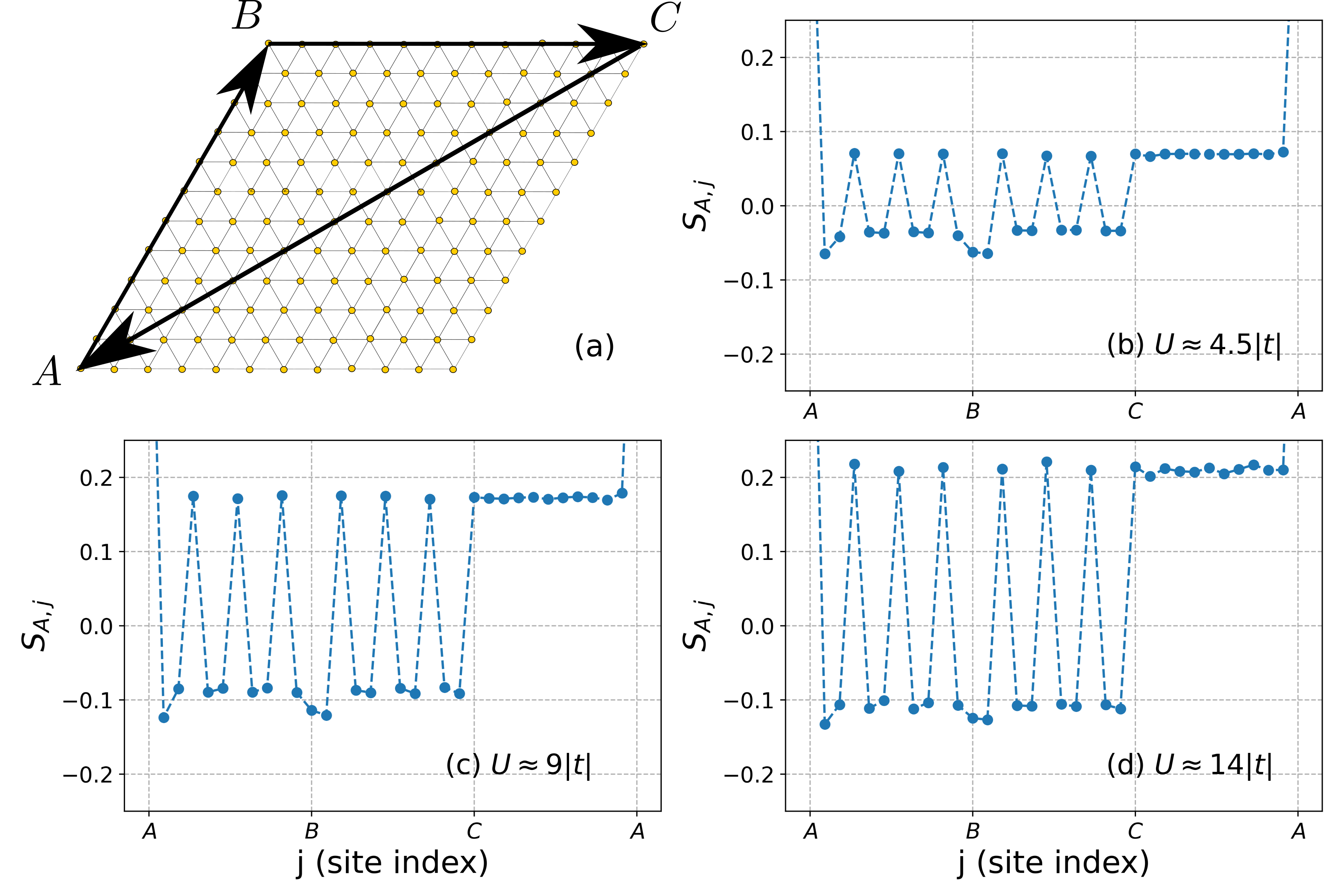}
    \caption{Real space spin-spin correlation functions $S_{A,j}$ for different values of the interaction amplitude $U$. (a) Sketch of the path along which the correlation functions have been collected; for example, in the A-B segment index $j$ refers to the lattice sites moving along the ${R}_{BA}$ vector. (b-d) Spatial dependence of the spin-spin correlation functions for representative values of $U$. }
    \label{fig:sisj}
\end{figure}

We start our analysis in the real space picture considering spin-spin correlation functions defined as
\begin{align}
    S_{i,j}\equiv\langle\hat{S}_i\cdot\hat{S}_j\rangle=\sum_{~\tau}^{\{x,y,z\}}\Big\langle\hat{S}^{~\tau}_i\hat{S}^{\tau}_j\Big\rangle,
    \label{eq:spinspinreal}
\end{align}
where  $\hat{S}_i^{\tau}$ are spin operators given in the second quantization language associated with direction $\tau\in\{x,y,z\}$ for electron residing at $i$-th site. 
In Fig.~\ref{fig:sisj} we present $S_{ij}$ correlation functions by setting the index $i=A$ [see Fig.\ref{fig:sisj}(a)] and vary $j$. The correlation functions $S_{A,j}$ exhibit a clearly visible -/-/+ pattern that repeats along the vectors $\mathbf{R}_{BA}$ and $\mathbf{R}_{BC}$. The absolute magnitudes of the negative values are about two times smaller than those of positive ones. For section $C-A$ we obtain only positive values with amplitude that do not show a clear decay with increasing distance between the lattice sites. These observations are consistent with the appearance of a $120^{\circ}$ AF spin order even below $U_{c1}$.  Furthermore, a separate analysis of $\langle\hat{S}_{A}^{z}\hat{S}_j^{z}\rangle$ has revealed its fast spatial decay and its absolute magnitude of about one order of magnitude smaller than $\langle\hat{S}_{A}^{x}\hat{S}_{j}^{x}+\hat{S}_{A}^{y}\hat{S}_{j}^{y}\rangle$ for $j\neq A$. Therefore, we conclude that AF develops in the $\mathbf{x}-\mathbf{y}$ plane. The absolute value of spin-spin correlations increases with $U$, however, for the entire range of onsite interactions considered, we report very weak or even no spatial decay beyond the next nearest neighbor.

To provide more direct evidence of the in-plane $120^{\circ}$ AF in the real-space picture, we have carried out semiclassical reasoning based on the mean values of correlation functions obtained for the spin ladder operators (see the Appendix for details). First, one may observe that when $\langle\hat{S}_{i}^{x}\hat{S}_{j}^{x}+\hat{S}_{i}^{y}\hat{S}_{j}^{y}\rangle \gg \langle\hat{S}_{i}^{z}\hat{S}_{j}^{z}\rangle$ (which holds in our case), the following is fulfilled,

\begin{align}
    \langle\hat{S}_{i}^{+}\hat{S}_{j}^{-}\rangle\approx\langle\hat{S}_i\cdot\hat{S}_j-i(\hat{S}_i\times\hat{S}_{j})\cdot\mathbf{z}\rangle,
    \label{eq:sspssm}
\end{align}
 since within the above assumption $\langle\hat{S}_i\cdot\hat{S}_{j}\rangle\approx \langle\hat{S}_{i}^{x}\hat{S}_{j}^{x}+\hat{S}_{i}^{y}\hat{S}_{j}^{y}\rangle$. Therefore, if we treat spins classically, the Eq. (\ref{eq:sspssm}) takes the form  
\begin{align}
    \langle\hat{S}_{i}^{+}\hat{S}_{j}^{-}\rangle\approx |\Vec{S}_i||\Vec{S}_j|\cos(\zeta_{i,j})-i|\vec{S}_i||\vec{S}_{j}|\sin(\zeta_{i,j}),
    \label{eq:spsm}
\end{align}
where $|\Vec{S}_{i}|$ ($|\Vec{S}_j|$) refers to the magnitude (length) of the \emph{classical} spin and $\zeta_{i,j}$ is the expectation value of the angle between the two spins. In this manner, one can extract the angle between the spins by knowing the complex values of $\langle\hat{S}_{i}^{+}\hat{S}_{j}^{-}\rangle$. Namely,
\begin{align}
    \zeta_{ij}\approx-\arg{ \langle\hat{S}_{i}^{+}\hat{S}_{j}^{-}\rangle}.
\end{align}
\begin{figure}
    \includegraphics[width=1.0\linewidth]{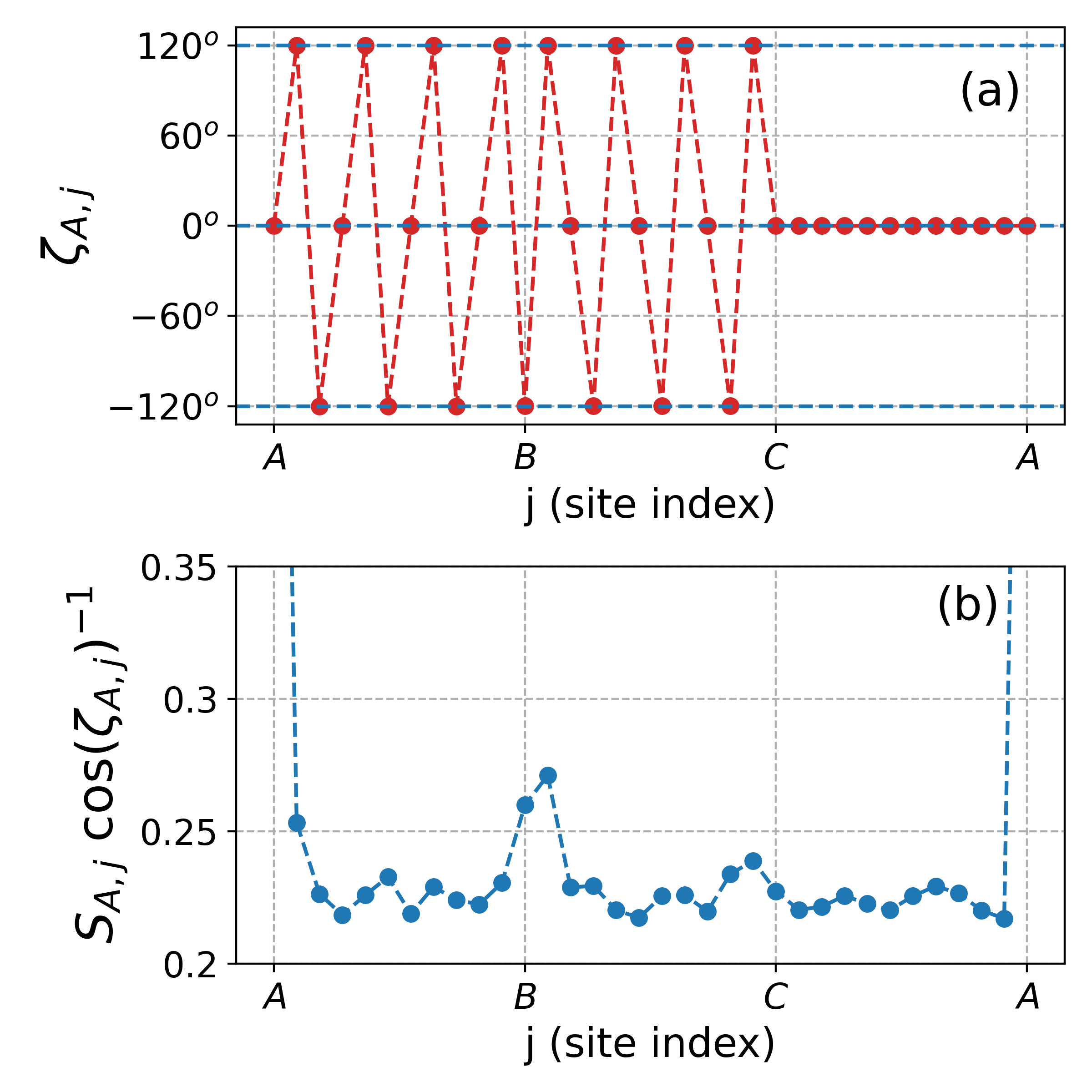}
    \caption{The value of $\zeta_{i,j}$ along the path defined in Fig.\ref{fig:sisj}(a). Spatial dependence of spin correlation function $S_{i,j}$ divided by the corresponding value of $\zeta_{i,j}$ collected for the sites included in the loop defined by vectors $\mathbf{R}_{BA}$, $\mathbf{R}_{CB}$ and $\mathbf{R}_{AC}$ (b).}
    \label{fig:zeta}
\end{figure}

In Fig.~\ref{fig:zeta}(a) we present $\zeta_{i,j}$ obtained for the path defined in Fig.~\ref{fig:sisj}(a). It clearly illustrates that the pattern -/-/+ refers to the angles $+120^{\circ}$/$-120^{\circ}$/$0^{\circ}$ confirming the presence of the $120^{\circ}$ AF order. This behavior has been observed for the entire range of $U$ under consideration. Moreover, the inspection of $S_{A,j}/\cos{(\zeta_{A,j})}$ presented in Fig.~\ref{fig:zeta}(b) shows that, disregarding discrepancies originating from numerical issues and statistical nature of the VMC method, correlations between spins are of similar amplitude when one removes the factor originating from their relative directions.

\begin{figure}
    \includegraphics[width=1.0\linewidth]{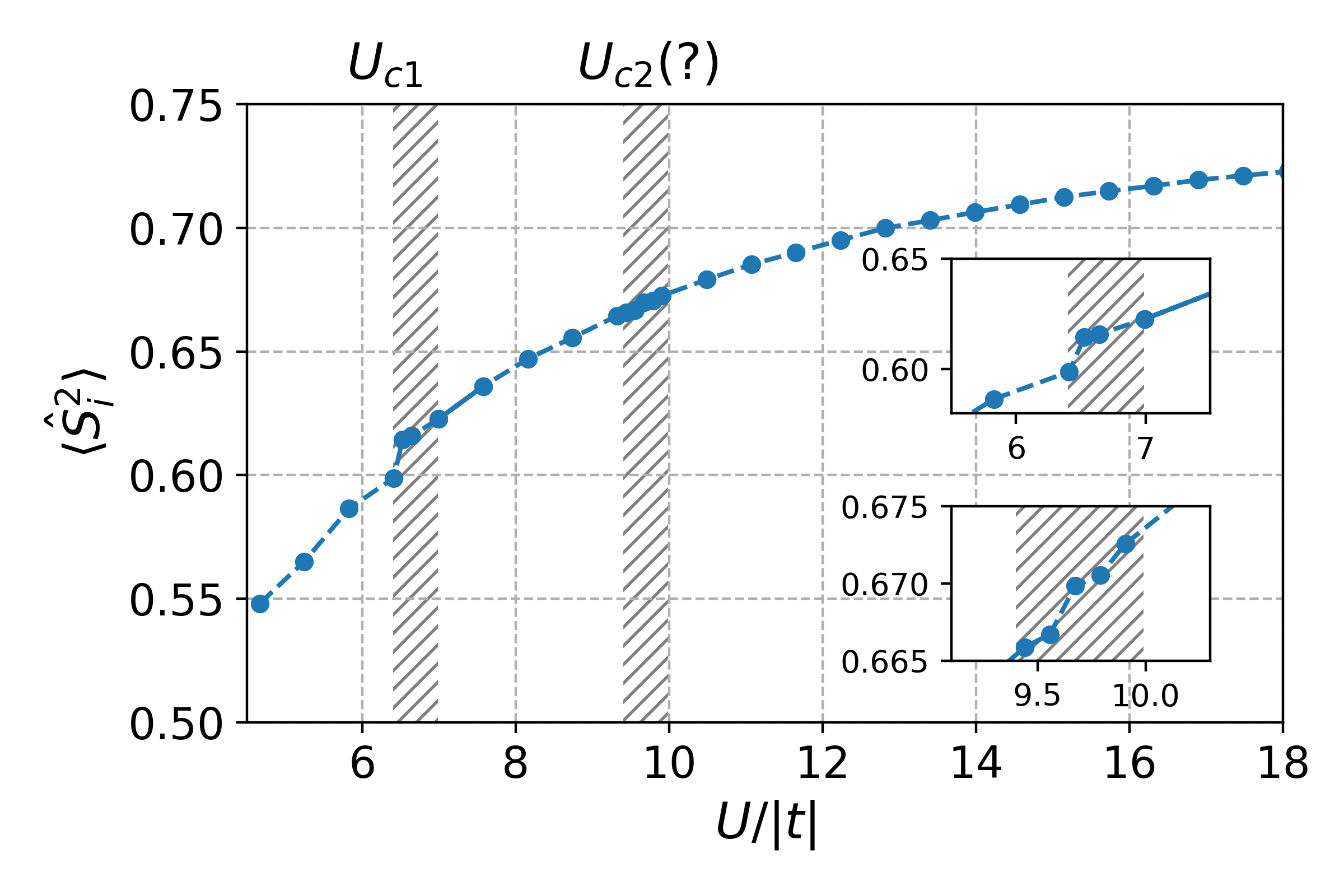}
    \caption{The expectation value of total spin squared per site as a function of $U$. The insets present the zoom of $S^{2}$ in the vicinity of the critical values of $U$.}
    \label{fig:s2}
\end{figure}

We complete our considerations in the real space by presenting the mean value of the total spin squared per site, ${S}^2=\frac{1}{L^2}\sum\langle\hat{S}_i^{2}\rangle$. Although we have not observed significant signatures of criticality at $U_{c1}$ and $U_{c2}$ in the above analysis, they are present by analyzing ${S}^2$ as a function of $U$, which is shown in Fig.~\ref{fig:s2}. Namely, at both $U_{c1}$ and $U_{c2}$ an abrupt change in the magnitude of squared spin appears. However, at $U_{c1}$ it is manifested more clearly than at $U_{c2}$ (see the insets in Fig.~\ref{fig:s2}). Apart from the two critical values of $U$, $S^2$ increases smoothly approaching the value of $S^2=3/4$, indicating an improvement in spin localization for $U\gg |t|$ as expected.

\begin{figure}
    \centering
    \includegraphics[width=1.0\linewidth]{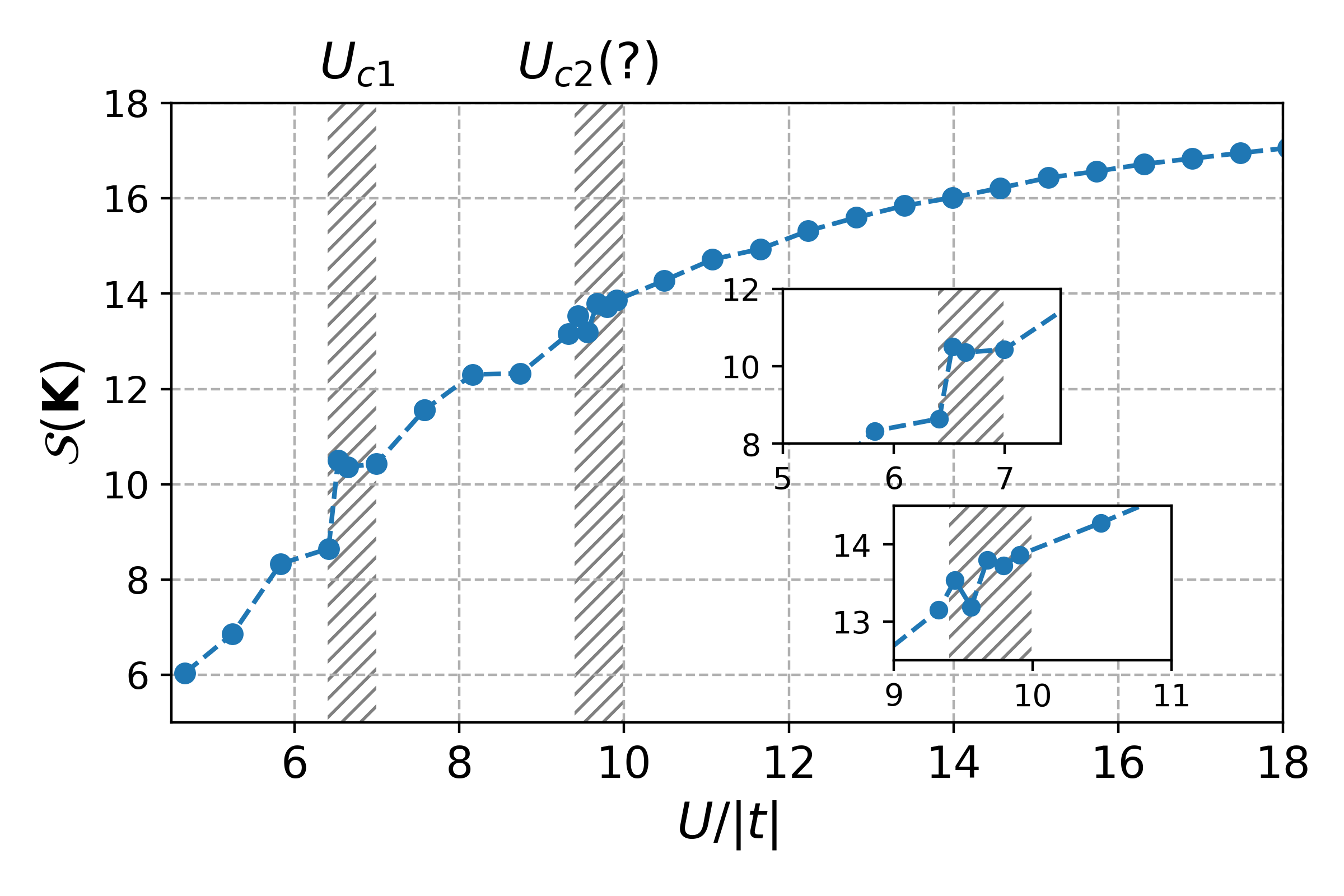}
    \caption{Momentum-resolved spin-spin correlation function at $\mathbf{q}=\mathbf{K}$, i.e., peak value of $\mathcal{S}(\mathbf{K})$. The insets show values in the vicinity of $U_{c1}$ and $U_{c2}$.}
    \label{fig:skpeak}
\end{figure}

In the following, we carry out the analysis of spin ordering in the momentum space. Here, we investigate the Fourier transform of the correlation functions defined in Eq.~(\ref{eq:spinspinreal}). Namely,
\begin{align}
    \mathcal{S}(\mathbf{q})=\frac{1}{L^2}\sum_{i,j}\text{e}^{i\mathbf{q}\cdot \mathbf{R}_{ij}}S_{i,j}.
\end{align}
\begin{figure}
    \centering
    \includegraphics[width=1.0\linewidth]{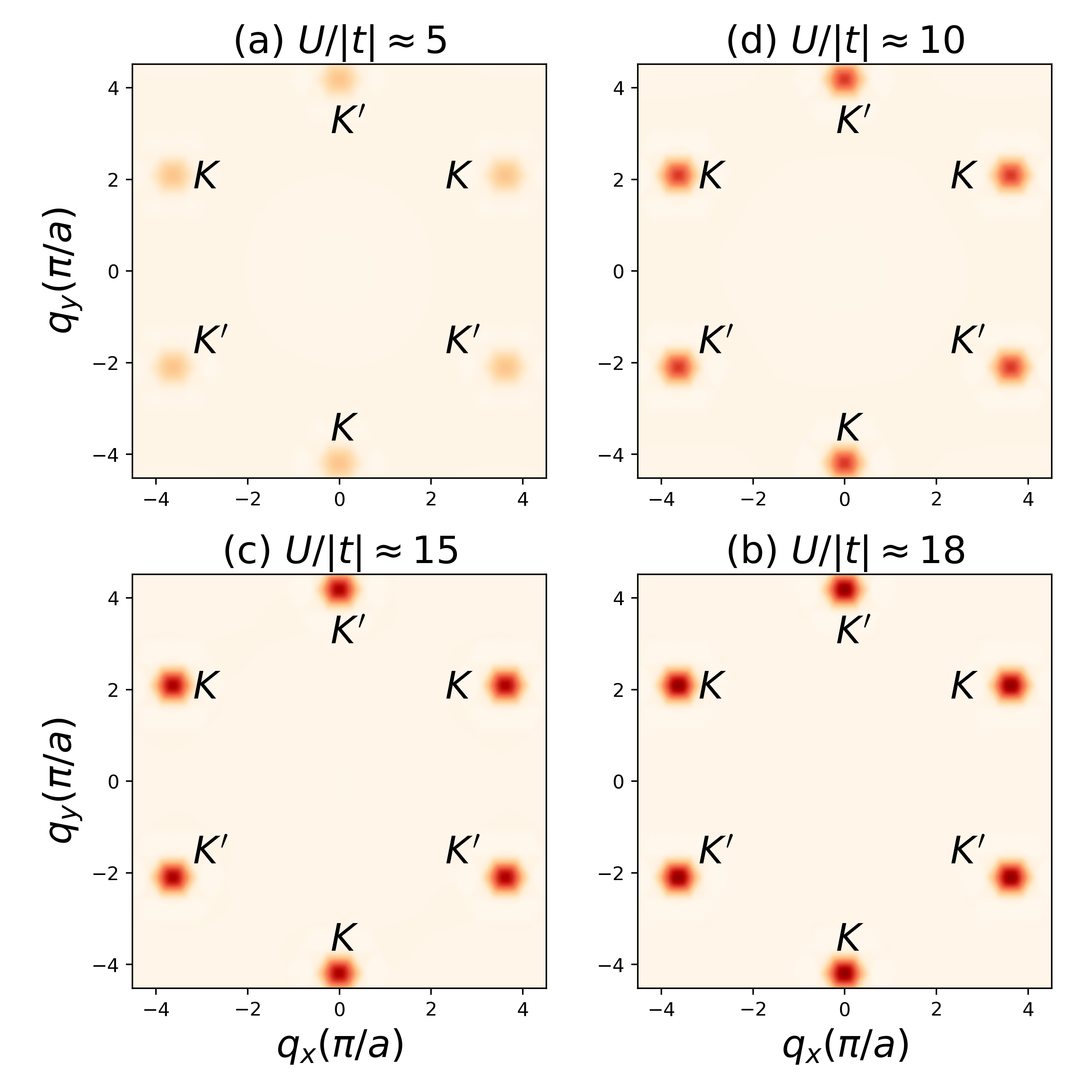}
    \caption{The maps presenting amplitudes of $\mathcal{S}(\mathbf{q})$ for the four representative values of $U$ (a-d). The increase in intensity of the red color indicates a higher value, and the color scale is the same for all the plots. Note, the we applied interpolation based smoothing for the presentation purposes. Peaks (red circles) are located exactly at $\mathbf{q}_{peak}\in\{\mathbf{K},\mathbf{K'}\}$.}
    \label{fig:skskmaps}
\end{figure}
In Fig.~\ref{fig:skskmaps}(a-d) we present the smoothed maps of $\mathcal{S}(\mathbf{q})$ for selected values of $U$. The peaks located exactly at $\mathbf{K}$ and $\mathbf{K}'$ reflect the appearance of the AF ordering that is consistent with the real space picture. 
The amplitudes of the peaks increase with $U$. In Fig.~\ref{fig:skpeak} we present $\mathcal{S}(\mathbf{K})=\mathcal{S}(\mathbf{K}')$ as a function of $U$. Again, the anomalous behavior of $\mathcal{S}(\mathbf{K})$ in the vicinity of $U_{c1}$ and a less pronounced discontinuity at $U_{c2}$ are both present.

Finally, we discuss the spin properties of the system in a strongly correlated state, i.e., for $U\approx16|t|$, as a function of the complex phase of the hoppings, $\phi$. As stated above, in the experimental situation, $\phi$ can be relatively easily tuned by the electric \emph{ displacement} field perpendicular to the twisted bilayer that originates from the top and bottom gates\cite{Haining2020,Wang2020}. Therefore, it is tempting to see how magnetically ordered states can be affected by changes in $\phi$. 
\begin{figure}
    \centering
    \includegraphics[width=1.0\linewidth]{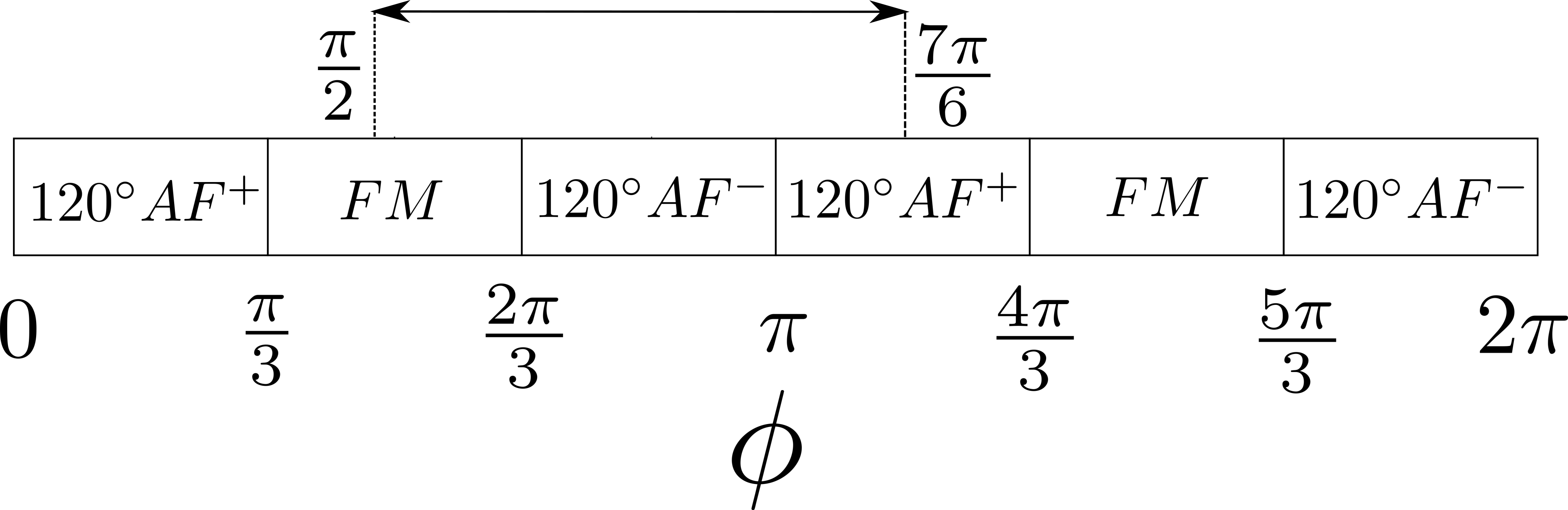}
    \caption{The sketch of magnetic phase diagram provided in [\onlinecite{Haining2020}] which can be deduced from the analysis of effective anisotropic Heisenberg model supplied with the Dzyaloshinskii-Moriya term. The plus(minus) sign in superscript of AF indicates clock(anticlock)-wise rotation of spin in the given frame of reference. In the upper part of the diagram, we mark the range $\phi\in(\pi/2,7\pi/6)$ which we have examined. }
    \label{fig:affm}
\end{figure}

As stated in Ref. \onlinecite{Haining2020} the  Hubbard model considered here, in the limit $U\gg|t|$, leads to an effective anisotropic Heisenberg model supplemented with the Dzyaloshinskii-Moriya term which in turn, based on semiclassical arguments, is believed to result in the magnetic phase diagram sketched in Fig.~\ref{fig:affm}. As one can see, the series of AF and FM phases is expected to appear in $\Delta \phi=\pi/3$ segments. Note that two degenerate phases denoted as AF$^{\pm}$ are predicted to exist within an approximate achievable experimental range of $\phi$ which is $(-2\pi/3,2\pi/3)$. These two states are distinguishable in the given frame of reference by considering the clock-wise rotation (AF$^{+}$) or anticlock(AF$^{-}$)-wise rotation of spins. 

\begin{figure}
    \centering
    \includegraphics[width=1.0\linewidth]{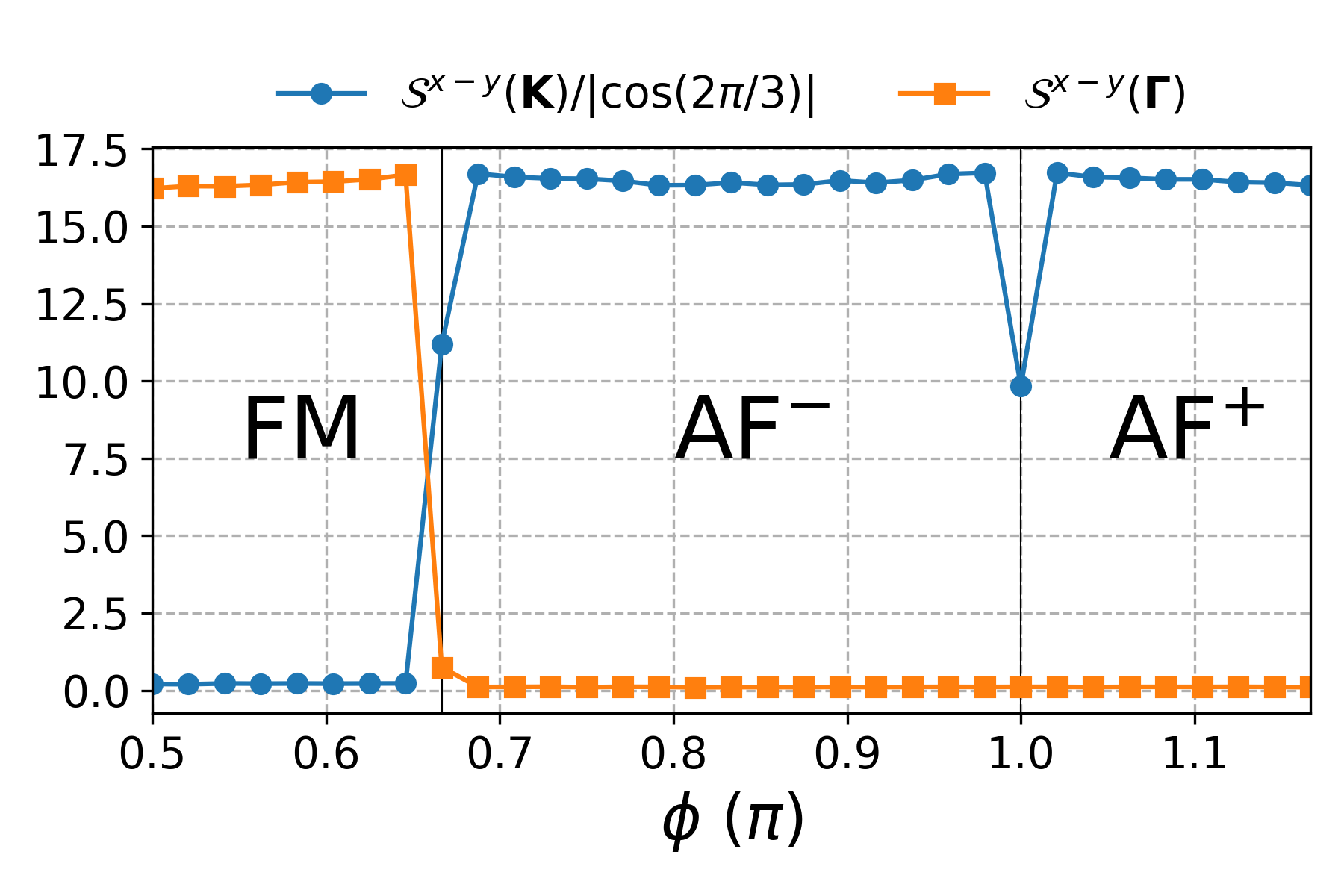}
    \caption{The peak values of $\mathcal{S}^{x-y}(\mathbf{q})$ functions (see main text) for the selected range of phase $\phi$. A solid vertical black lines indicate $\phi=2\pi/3$ and $\phi=\pi$. }
    \label{fig:fmaf_sxy}
\end{figure}

\begin{figure}
    \centering
    \includegraphics[width=1.0\linewidth]{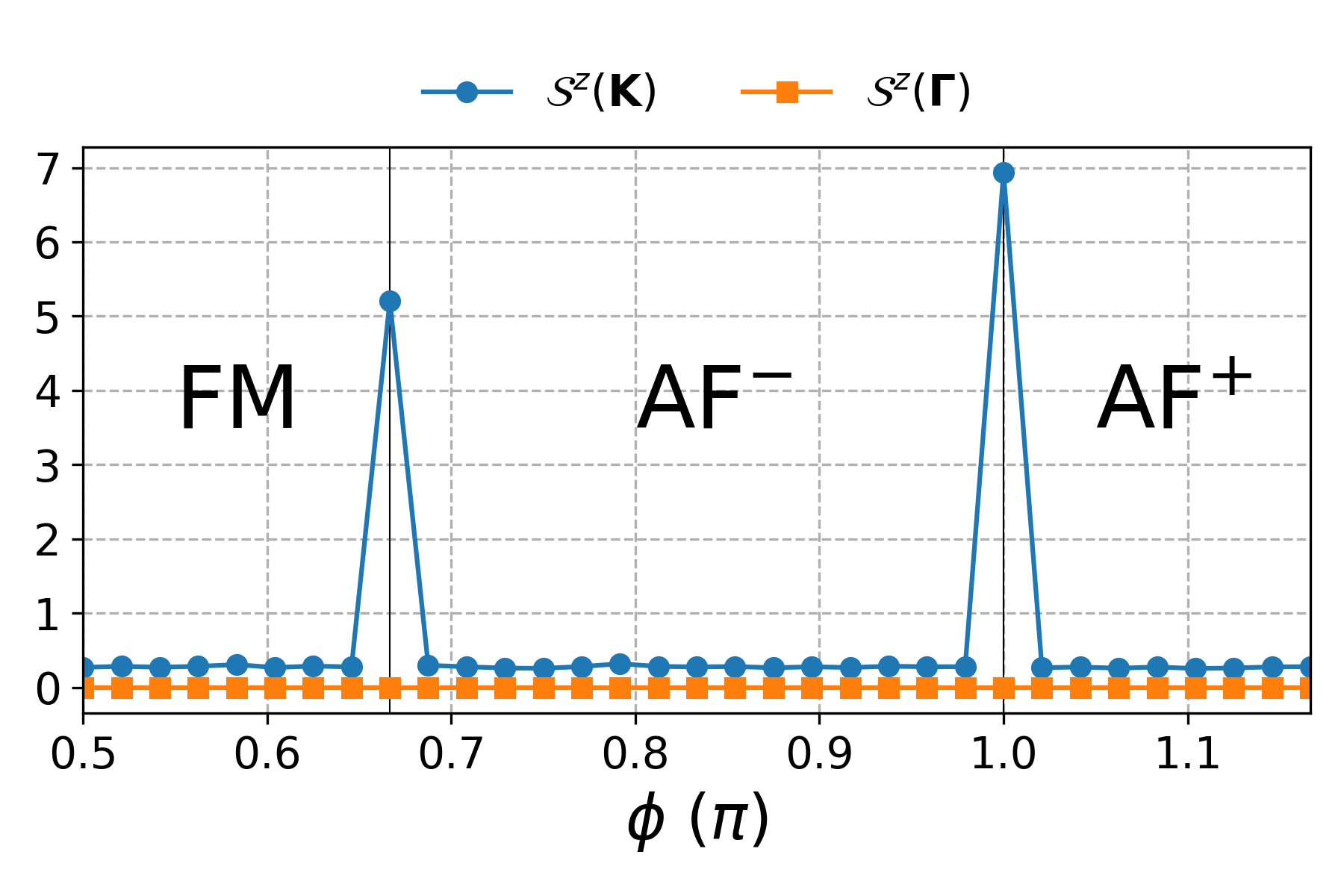}
    \caption{Fourier transform of the $\mathbf{z}$-component of spin-spin correlation functions  at $\mathbf{q}\in\{\mathbf{\Gamma},\mathbf{K}\}$ for the considered range of phase $\phi$. Vertical solid black line indicates $\phi=2\pi/3$ and $\phi=\pi$.}
    \label{fig:fmaf_sz}
\end{figure}

In our analysis, we focus on $\phi\in(\pi/2,7\pi/6)$ and investigate separately the in-plane and out-of-plane  spin-spin correlation functions in the momentum space,
\begin{align}
\mathcal{S}^{x-y}(\mathbf{q})\equiv\frac{1}{L^2}\sum_{i,j}\sum_{\tau\in\{x,y\}}\text{e}^{i\mathbf{q}\cdot\mathbf{R_{ij}}}\langle\hat{S}_i^{\tau}\hat{S}_j^{\tau}\rangle,
\end{align}
and,
\begin{align}
 \mathcal{S}^{z}(\mathbf{q})  \equiv\frac{1}{L^2}\sum_{i,j}\text{e}^{i\mathbf{q}\cdot\mathbf{R_{ij}}}\langle\hat{S}_i^{z}\hat{S}_j^{z}\rangle,
\end{align}
 respectively. Our results show peaks in the mentioned correlation functions only at $\mathbf{q}=\mathbf{K}(\mathbf{K}')$ or $\mathbf{q}=\mathbf{\Gamma}$ depending on phase $\phi$. In Fig.~\ref{fig:fmaf_sxy} we present $\mathcal{S}^{x-y}(\mathbf{\Gamma})$  and $\mathcal{S}^{x-y}(\mathbf{K})/|\cos(2\pi/3)|$, as a function of $\phi$. It is clearly visible that the system is in FM state when $\phi<2\pi/3$ since the peaks in this range are located at $\mathbf{\Gamma}$ and the amplitude of $\mathcal{S}^{x-y}(\mathbf{K})$  is residual. The in-plane character of FM ordering is confirmed by the observation that $\mathcal{S}^{z}(\mathbf{\Gamma})$ remains nearly zero in the whole range of the considered $\phi$, as shown from Fig.~\ref{fig:fmaf_sz}. 
For $\phi>2\pi/3$ the AF state is stable since $\mathcal{S}^{x-y}(\mathbf{\Gamma})\approx 0$ and the magnitude of $\mathcal{S}^{x-y}(\mathbf{K})$ is the same as $\mathcal{S}^{x-y}(\mathbf{\Gamma})$ for $\phi<2\pi/3$. The values of $\mathcal{S}^z(\mathbf{K})$ are an order of magnitude smaller than $\mathcal{S}^{x-y}(\mathbf{\Gamma})$ $\phi<2\pi/3$ or $\mathcal{S}^{x-y}(\mathbf{K})$ for $\phi>2\pi/3$. 

Interestingly, at $\phi=2\pi/3$
we find that $\mathcal{S}^z(\mathbf{K})\approx\mathcal{S}^{x-y}(\mathbf{K})/2$, which indicates an out-of-plane AF phase in between the in-plane FM and AF stability regions. Therefore, it seems that at $\phi=2\pi/3$ the system behaves similarly to the case when $\phi=\pi$, since then the Hubbard model at $U\gg|t|$ maps to the isotropic Heisenberg model (without the Dzyaloshinksii-Moriya term) for which 120$^{\circ}$ out-of-plane AF emerges\cite{White2007}. The formation of this state, explicitly for the \emph{standard} (with real valued, spin-independent hopping terms) Hubbard model  on the triangular lattice has been recently reported by e.g. Chen et al.\cite{Chen2022}. Finally, at $\phi=\pi$ we also find substantial contribution to the AF order coming from the $z$-direction since the magnitude of $\mathcal{S}^z(\mathbf{K})$ is even greater than for $\phi=2\pi/3$ (for $\mathcal{S}^{x-y}(\mathbf{K})$ the opposite holds). Therefore, at $\phi=\pi$ we find spin order signatures that can be considered as those emerging from the \emph{standard} Hubbard model on a triangular lattice at half filling.
Further increase in $\phi$ results in the formation of  antiferromagnetic state characterized by the opposite chirality when compared to that referring to $2\pi/3<\phi<\pi$  - in agreement with the predictions presented in [\onlinecite{Haining2020}].

\section{Summary and conclusions}

We have analyzed the Hubbard model on the triangular lattice which is believed to describe fundamental electronic features of WSe${_2}$ twisted homobilayer at half filling. Our comprehensive analysis based on the VMC approach reveals the emergence of the Mott insulating state at $U_{c1}\approx6.5|t|\div7|t|$ as can be extracted from several indicators considered here which involve correlation functions resolved both in real and momentum spaces. Since free particle terms effectively generate valley-dependent spin-splitting, the model considered here can be regarded as a Hubbard-type Hamiltonian with reduced degrees of freedom\cite{Haining2020}.

The critical $U$ for the Mott gap opening obtained here is smaller than the one determined earlier ($U\approx8|t|$) for the case of standard Hubbard model on a triangular lattice with real hopping parameters (see e.g.  [\onlinecite{Arovas2022},\onlinecite{Zampronio2023}] and the references therein). Furthermore, we observe some anomalies at $U_{c2}\approx9.5|t|\div10|t|$ by inspecting both the density-density and spin-spin correlation functions. However, the tendency towards formation of the long-ranged  $120^{\circ}$ AF order is present even below $U_{c1}$ and is enhanced with increasing $U$. Therefore, we do not find signatures which would allow us to clearly distinguish the phases $U_{c1}\lesssim U \lesssim U_{c2}$ and $U\gtrsim U_{c2}$ in addition to pointing out the anomalies themselves. 

It can be speculated that the robustness of AF (lack of evidence for the long-range spatial decay) even below $U_{c1}$ originates from the valley dependent spin-splitting which drives the system towards AF correlations. However, for the \emph{standard} Hubbard model (without the spin-splitting feature), peaks in spin structure factor $\mathcal{S}(\mathbf{q})$ have also been reported in the metallic regime but with clearly visible spatial decay in the spin-spin correlation functions\cite{Shirakawa}. Also, incorporation of the  AF order into the $x-y$ plane (in-plane AF) which appears in the insulating state can be regarded as a clear differentiation between the \emph{ standard} Hubbard model and the one considered here.

Finally, it should be noted that according to previous reports the complex phase of the hoppings which introduce the valley-dependent spin-splitting can be tuned with the use of the displacment field. Therefore, we have analyzed if the modification of phase $\phi$ may result in switching between the AF and the FM states as proposed in Ref. \onlinecite{Haining2020} based on a semiclassical reasoning concerning the anisotropic Heisenberg model with a Dzyaloshinskii-Moriya term. We have confirmed that the decrease of $\phi$ below $2\pi/3$ switches the system from in-plane AF order to the in-plane ferromagnetic order. It is also worth mentioning that at $\phi=2\pi/3$ system exhibits a property similar to that typical for the Hubbard model without a spin-splitted band, namely, AF order with non-zero component related to $\mathbf{z}$ direction. The possibility of switching between magnetic states (AF$\rightarrow$FM) by using the electric field could be interesting in the view of possible applications in modern electronics. To make it achievable the regime of displacement fields corresponding to the considered range of $\phi$ would have to be reached experimentally.

The data behind all the figures are available in the open repository\cite{andrzej_biborski_zenodo}.

\section{Acknowledgement}
This research was founded by National Science Centre, Poland (NCN) according to decision 2021/42/E/ST3/00128. 
\vspace{1cm}
\appendix*
\label{appendix}
\section{Approach for retriving in-plane angle $\zeta_{i,j}$ 
from spin ladder operators correlation function. }
The expectation value of angle between spins, $\zeta_{i,j}$, treated as classical vectors can be extracted in our approach in the following manner. It should be noted that for the obtained in-plane AF state the $z$-th component of the spin-spin correlations is negligible when compared to those related $\mathbf{x}$ and $\mathbf{y}$ directions. Thus, 
\begin{align}
\begin{split}
    \langle \hat{S}_i\cdot\hat{S}_j\rangle\approx\langle \hat{S}_i^x\hat{S}_j^x+\hat{S}_i^y\hat{S}_j^y\rangle=\\=\Big\langle\frac{1}{2}(\hat{S}_i^+\hat{S}_j^-+\hat{S}_i^-\hat{S}_j^+)\Big\rangle=\Re\langle\hat{S}_i^{+}\hat{S}_j^{-}\rangle,
\end{split}
\end{align}
since $\hat{S}_{i}^{+}\hat{S}_j^{-}=\big(\hat{S}_{i}^{-}\hat{S}_j^{+}\big)^{\dagger}$. On the other hand, we have,

\begin{align}
\begin{split}
\langle\hat{S}_i^{+}\hat{S}_j^{-}\rangle=\big\langle(\hat{S}_i^{x}+i\hat{S}_i^{y})(\hat{S}_j^{x}-i\hat{S}_j^{y})\big\rangle=\\
=\langle\hat{S}_i^{x}\hat{S}_j^{x}+\hat{S}_i^{y}\hat{S}_j^{y}-i\hat{S}_i^{x}\hat{S}_j^{y}+i\hat{S}_i^{y}\hat{S}_j^{x}\rangle\approx\\
\approx\big\langle\hat{S}_i\cdot\hat{S}_j-i(\hat{S}_i\times\hat{S}_j)\cdot\mathbf{z}\big\rangle.
\end{split}
\end{align}
Note that
\begin{align}
\langle\hat{S}_{i}^{+}\hat{S}_j^{-}\rangle=\langle\hat{S}_{i}^{-}\hat{S}_j^{+}\rangle^{*},    
\end{align}
hence both $\langle\hat{S}_i\cdot\hat{S}_j\rangle$ and $\big\langle(\hat{S}_i\times\hat{S}_j)\cdot\mathbf{z}\big\rangle$ is real. Therefore, we may attribute the classical inner and cross products between \emph{classical} spins $\Vec{S}_i$ and $\Vec{S}_j$ to the former and latter averages, respectively,
\begin{align}
\begin{split}
        \langle\hat{S}_i^{+}\hat{S}_{j}^{-}\rangle\equiv|\Vec{S}_{i}||\Vec{S}_j|\cos{\zeta_{i,j}}-i|\Vec{S}_{i}||\Vec{S}_j|\sin(\zeta_{i,j})=\\=|\Vec{S}_{i}||\Vec{S}_j|\text{e}^{-i\zeta_{i,j}}.
\end{split}
\label{eq:zeta}
\end{align}

Finally, by extracting the value of $\langle\hat{S}_i^{+}\hat{S}_{j}^{-}\rangle$ from our VMC calculation scheme, we can calculate the angle $\zeta_{i,j}$ in the following manner
\begin{align}
    \zeta_{i,j}=-\arg \langle\hat{S}_i^{+}\hat{S}_{j}^{-}\rangle.
\end{align}

\bibliography{refs.bib}

\end{document}